\documentclass[sigconf]{acmart}

\usepackage{booktabs}
\usepackage{tabularx}
\usepackage{array}

\usepackage{hyperref}

\usepackage{multirow}
\usepackage{graphicx}
\usepackage{array}

\newcommand{\rotgroup}[1]{
\rotatebox[origin=c]{90}{
\parbox[c]{2.55cm}{\centering\bfseries #1}
}
}

\AtBeginDocument{%
  }

\setcopyright{acmlicensed}
\copyrightyear{2026}
\acmYear{2026}
\setcopyright{cc}
\setcctype{by}
\acmConference[ICS Workshops '26]{2026 International Conference on Supercomputing Workshops}{July 06--09, 2026}{Belfast, United Kingdom}
\acmBooktitle{2026 International Conference on Supercomputing Workshops (ICS Workshops '26), July 06--09, 2026, Belfast, United Kingdom}
\acmDOI{10.1145/3774895.3815552}
\acmISBN{979-8-4007-2300-1/2026/07}

\setcopyright{none}
\renewcommand\footnotetextcopyrightpermission[1]{}

\begin{document}

\title{When Noisy Quantum Order Finding Remains Recoverable for Shor’s Algorithm}

\author{Qingxin Yang}
\affiliation{%
  \institution{KTH Royal Institute of Technology}
  \city{Stockholm}
\country{Sweden}}
\email{qingxiny@kth.se}

\author{Stefano Markidis}
\affiliation{%
  \institution{KTH Royal Institute of Technology}
  \city{Stockholm}
  \country{Sweden}}
  \email{markidis@kth.se}

\renewcommand{\shortauthors}{Yang and Markidis}

\begin{abstract}
Order finding is the core subroutine of Shor’s algorithm. On NISQ hardware, its phase estimation output distributions are often distorted by noise, making correct order recovery difficult. We study \emph{recoverability} in noisy order finding: given a measured precision-register distribution, when does standard classical post-processing still return the true order? We analyze 680 distributions collected from IBM quantum systems across multiple problem instances and circuit settings. For each distribution, we apply continued-fraction post-processing with modular verification and define recoverability as a binary outcome according to whether the recovered order equals the true one. We then characterize each distribution using four features: autocorrelation peak strength, normalized entropy, dominant verified mass fraction, and verified margin fraction. We evaluate these quantities using marginal feature comparisons, single-feature AUROC analysis, and multivariate tree-based classifiers. We further use random-forest permutation importance to assess which quantities contribute distinct predictive information once the other features are known. To make the resulting classification behavior interpretable, we also train a decision tree that exposes threshold rules for recoverable and non-recoverable distributions. We find that recoverability is strongly associated with both residual comb-like structure in the measured distribution and the way verified probability mass is organized across candidate denominators. The dominant verified mass fraction is the strongest single-feature indicator of recoverability, and the tree-based analysis shows that it also provides the primary split in an interpretable threshold description. Some highly distorted distributions remain recoverable when one verified denominator dominates the post-processing mass, while some visibly structured distributions fail because classical post-processing favors an incorrect verified denominator.
\end{abstract}



\maketitle

\section{Introduction}
\begin{figure*}[t]
    \centering
    \includegraphics[width=\linewidth]{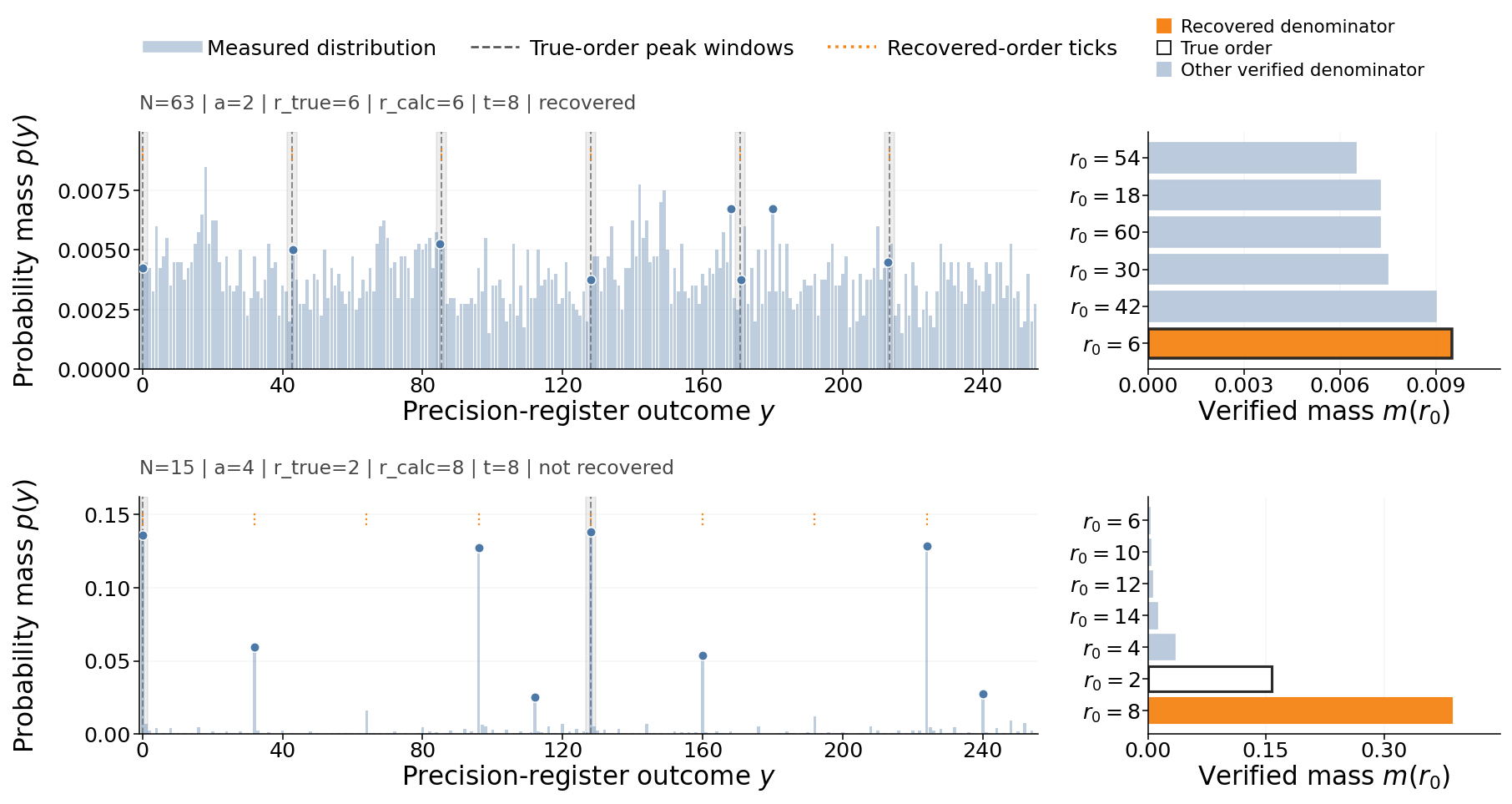}    
    \caption{Top: Recoverability despite weak global histogram structure. Although the measured distribution is diffuse and the global structure metrics are weak, the verified candidate comparison still selects the true period, \(r=6\), yielding recoverability. Bottom: No recoverability despite visible peak structure. The verified candidate comparison favors an incorrect period, \(r=8\), over the true period, \(r=2\), and recoverability fails.}
    \label{fig:case1_case2}
\end{figure*}

Quantum order finding is the main quantum subroutine in Shor's factoring algorithm and, more broadly, a fundamental application of quantum phase estimation~\cite{shor1997polynomial,kitaev1995quantum}. Given coprime integers $a$ and $N$, the task is to determine the multiplicative order $r$ of $a$ modulo $N$, i.e. $a^r \equiv 1 \pmod N$, where $r$ is the smallest positive integer satisfying this relation. In the standard quantum routines, order information is encoded in the measurement distribution of a $t$-qubit phase-estimation (also called \emph{precision}) register. Writing $Q=2^t$ and $y\in\{0,\dots,Q-1\}$ for the measured precision register value, the output is the distribution $p(y)=\Pr(Y=y)$. In the noiseless ideal setting, this distribution has a characteristic peak-comb structure (see, for instance, the bottom left panel of Figure~\ref{fig:case1_case2}). Its probability mass is concentrated near the locations $y\approx Qs/r$, corresponding to phases $s/r$ for $s=0,\dots,r-1$.

However, on Noisy Intermediate-Scale Quantum (NISQ) hardware~\cite{preskill2018quantum}, measured precision-register distributions are distorted relative to this ideal peak structure. Peaks may broaden, shift, merge, or partially flatten, and the resulting distribution can appear far from the noiseless output. An example of such distortion is presented in the top left panel of Figure~\ref{fig:case1_case2}. However, such visual distortion does not by itself imply algorithmic failure. In fact, in a nontrivial subset of cases, classical post-processing applied to a noisy histogram still recovers the correct order. 

Figure~\ref{fig:case1_case2} provides a motivating example for this work. The top row shows a run whose measured precision register distribution is highly diffuse and visually far from an ideal comb. Nevertheless, the right-hand panel shows that the verified-denominator masses \(m(r_0)\) are maximized at the true order \(r=6\), so the run is recoverable. The bottom row shows the opposite phenomenon. The measured distribution shows clear peaks and appears less degraded. Yet, the verified-denominator masses favor an incorrect denominator \(r_{\mathrm{calc}}=8\) over the true order \(r=2\), so recoverability fails.  
The distributions on the left are informative. However, they are not decisive. What determines success is how the measured probability mass is redistributed by the actual classical post-processing map on the right.

The objective of this work is to study noisy quantum order finding through an interpretable statistical analysis of measured precision-register distributions. In this work, we ask which observable properties of noisy output distributions are most strongly associated with successful order recovery. The key question is not only how far a measured distribution is from an ideal peak-comb distribution, but also whether the remaining probability mass is organized in a way that still supports correct order recovery after standard classical post-processing. Such an analysis is also potentially useful for application-specific quantum error mitigation, since it can identify which distribution-level properties should be preserved or restored in order to improve recoverability.

In this work, we use the notion of \emph{recoverability}. For each measured distribution, we apply the standard continued-fraction post-processing routine, retain only candidate denominators that satisfy modular verification, and aggregate probability mass over the resulting verified denominators. The inferred denominator \(r_{\mathrm{calc}}\) is defined as the verified denominator with the largest aggregated mass. A run is then called recoverable if \(r_{\mathrm{calc}}=r\), where \(r\) is the true order. 

This definition also explains why the right-hand panels of Figure~\ref{fig:case1_case2} are essential. 
They make it explicit that recoverability depends on mass competition among verified candidate denominators, not only on a visually comb-like distribution. Following this reasoning, the quantities used in this work are designed to separate these two effects. We first use two global distribution descriptors. These are autocorrelation peak strength \(A_{\mathrm{peak}}\), which measures the repeated-spacing structure, and normalized entropy \(H_{\mathrm{norm}}\), which measures flatness. We also use two post-processing-aware but order-unknown quantities derived from the verified candidate masses. These are the dominant verified mass fraction \(M_{1,\mathrm{frac}}\), which measures how strongly verified mass concentrates on the leading candidate denominator, and verified margin fraction \(\Delta_{\mathrm{ver,frac}}\), which measures how far that leading candidate is separated from its nearest verified competitor. 

We evaluate this framework on 680 precision-register distributions collected from IBM quantum systems using a structured family of Shor-style order-finding circuits.  The benchmark covers multiple problem instances, precision-register sizes, and approximate-QFT settings. We find that recoverability is governed jointly by residual comb-like structure in the measured histogram and by the redistribution of verified mass across candidate denominators. Recoverable runs tend to have stronger autocorrelation peaks, lower normalized entropy, larger dominant verified mass fractions, and larger verified margins. 

This work focuses on interpretability at the predictive level. Single-feature AUROC analysis shows that all four quantities are informative, with post-processing-aware quantities ranking highest. In addition, simple tree-based models already lead to explicit threshold-based descriptions of recoverability.

The main contributions of this work are as follows:
\begin{itemize}
 \item We investigate, with a data-driven approach, which properties of noisy precision-register distributions are most strongly associated with recoverability, distinguishing between global distribution structure and post-processing-aware verified-candidate quantities.
 \item We show that algorithm-aware quantities derived from the verified probability masses provide the strongest indicators of recoverability. Global distribution descriptors such as autocorrelation and entropy provide complementary information.
 \item We present a hardware study on 680 measured precision-register distributions from IBM quantum systems, identifying statistical trends that separate recoverable from non-recoverable runs.
 \item We show that simple interpretable machine-learning models, in particular decision trees and random forests, give a predictive account of recoverability and help reveal which quantities determine recoverability success.
\end{itemize}

\subsection{Related work}

\noindent \textbf{Quantum Noise and Quantum Error Mitigation}. Prior work on noise in near-term quantum processors has primarily focused on physical error mechanisms, hardware constraints, and circuit-level remedies. Relevant limitations include imperfect gates, decoherence, readout errors, crosstalk, and constraints imposed by limited qubit connectivity, which can increase routing overhead and circuit depth. For phase-estimation-based algorithms, several approaches reduce circuit depth or resource requirements through semiclassical or approximate quantum Fourier transforms, qubit recycling, and scalable arithmetic constructions for Shor-type order finding~\cite{griffiths1996semiclassical,barenco1996approximate,coppersmith2002approximate,martinlopez2012experimental,monz2016realization}. Fault-tolerant resource estimates for cryptanalytic applications provide a complementary perspective. Gidney and Eker{\aa} analyzed large-scale RSA factoring with noisy physical qubits and surface-code protection, emphasizing the role of arithmetic design, repeated attempts, spacetime layout, and noise assumptions in realistic implementations of Shor-type algorithms~\cite{gidney2021factor}. These methods aim to improve phase estimation by reducing the number of quantum operations, the number of required qubits, or the exposure of the quantum circuit to noise. Another major direction is quantum error mitigation, which seeks to reduce the effect of noise on quantities estimated from noisy quantum circuits without requiring full fault tolerance. Representative techniques include zero-noise extrapolation and probabilistic error cancellation~\cite{temme2017error}, readout-error mitigation~\cite{bravyi2021mitigating}, and symmetry-based methods~\cite{bonetmonroig2018low}, with broader surveys given in Refs.~\cite{endo2021hybrid,cai2023quantum}. In the longer term, quantum error correction provides a systematic route to suppressing logical errors, with surface-code architectures among the leading candidates for scalable fault-tolerant quantum computation~\cite{fowler2012surface}. These approaches address noise by modifying circuit executions, correcting measurement statistics, statistically mitigating observables, or actively protecting encoded quantum information. A related line of work studies noisy quantum devices through characterization, verification, validation, and data-driven modeling. Randomized benchmarking and gate-set tomography estimate quantities such as average error rates and self-consistent gate models~\cite{magesan2011scalable, merkel2013self, nielsen2021gate}. Machine-learning methods have also been used for quantum error-correction decoding, including neural-network decoders for topological and surface codes and recent learned decoders trained on quantum processor data~\cite{torlai2017neural,varsamopoulos2018decoding,bausch2024learning}. Learned models have further been explored for error mitigation and device-adaptive correction of noisy expectation values~\cite{strikis2021learning,czarnik2021error,liao2024machine}. \\

\noindent \textbf{Algorithmic Post-Processing Methods}. Most closely related to our work are algorithmic post-processing methods that extract classical information from quantum measurement data, including both idealized post-processing analyses and explicitly noisy settings. In order finding, Eker{\aa} analyzed the success probability of Shor's order-finding subroutine and showed that limited classical searches can improve recovery without rerunning the quantum circuit~\cite{ekera2022success}. The same author also studied how an order-finding output can be used to completely factor an integer in a single run, emphasizing the role of classical post-processing after the quantum sampling step~\cite{ekera2021complete}. For noisy Simon period finding, May, Schlieper, and Schwinger showed that erroneous period-finding samples can remain useful after classical smoothing and reduction to a noisy learning problem~\cite{may2021noisy}. For phase estimation, O'Brien, Tarasinski, and Terhal studied Bayesian and time-series post-processing for extracting eigenvalues from low-depth noisy experiments~\cite{obrien2019quantum}, while Bayesian phase-estimation methods infer phases and uncertainty from experimental outcomes~\cite{wiebe2016efficient}. Related hybrid approaches include maximum-likelihood amplitude estimation, iterative quantum amplitude estimation, and curve-fitted quantum phase estimation~\cite{tanaka2021amplitude,grinko2021iterative,lim2024curve}. Our work follows this algorithmic post-processing perspective, but studies a different object. We analyze when measured precision-register distributions lead to successful order recovery and focus on distribution-level and post-processing-aware indicators of success.

\section{Preliminaries}
This section defines the recoverability criterion used throughout the paper. The goal is to convert a noisy precision-register distribution into a binary algorithmic outcome. Let us fix an order-finding instance \((N,a)\), with \(a\in(\mathbb Z/N\mathbb Z)^\times\), and let $r=\operatorname{ord}_N(a)$
denote the true multiplicative order of \(a\) modulo \(N\). 
Let \(p(y)\) be the noisy probability distribution on precision-register outcomes $y\in\{0,\ldots,Q-1\}$. For each outcome \(y\), the classical post-processing routine approximates \(y/Q\) by a continued-fraction convergent \(s/r_0(y)\), with denominator \(r_0(y)\le N\), as in Shor's order-finding algorithm~\cite{shor1997polynomial,nielsen2010quantum}. A candidate denominator is retained only if it passes modular verification: $a^{r_0(y)}\equiv 1 \pmod N$.
The main quantity in our analysis is the probability mass assigned to each verified candidate denominator. For each denominator \(r_0\), define its verified probability mass by
\begin{equation}
m(r_0) = \sum_{\substack{y:\, r_0(y)=r_0\\ a^{r_0}\equiv 1 \!\!\!\pmod N}} p(y).
\end{equation}
The total verified mass is $M_{\mathrm{ver}}=\sum_{r_0}m(r_0)$. If \(M_{\mathrm{ver}}=0\), then no positive-probability outcome is mapped to a denominator satisfying modular verification, and the distribution is declared non-recoverable. 
Otherwise, we define the inferred denominator as the verified denominator with the largest aggregated mass:
\begin{equation}
r_{\mathrm{calc}}\in \arg\max_{r_0} m(r_0),
\end{equation}
with ties resolved by a fixed deterministic rule. We say that the noisy distribution is \emph{recoverable} if $r_{\mathrm{calc}}=r$.

\section{Methods} 
This section describes the dataset and the distribution quantities used to analyze recoverability in noisy quantum order finding. \\

\begin{table}[t]
\centering
\setlength{\tabcolsep}{3pt}
\renewcommand{\arraystretch}{1.08}
\begin{tabularx}{\columnwidth}{>{\raggedright\arraybackslash}p{0.42\columnwidth}X}
\toprule
\textbf{Statistic} & \textbf{Value} \\
\midrule
Total runs & 680 \\
Recoverable runs & 310 \; (45.6\%) \\
Non-recoverable runs & 370 \; (54.4\%) \\
Backends & \texttt{ibm\_kingston} (390), \texttt{ibm\_boston} (170), \texttt{ibm\_torino} (80), \texttt{ibm\_pittsburgh} (30), \texttt{ibm\_miami} (10) \\
Instances $N$ & $\{3,7,15,31,63,127\}$ \\
Bases $a$ & $\{2,4,8,16\}$ \\
Precision size $t$ & 8 (595 runs), 10 (85 runs) \\
Approximate-QFT degree & 0 (80), 1 (85), 2 (515) \\
2Q-gate depth & min 96, median 1279, max 1997 \\
2Q-gate count & min 118, median 1654, max 2681 \\
\bottomrule
\end{tabularx}
\caption{Summary of the hardware order-finding dataset used in this work.}
\label{tab:dataset_summary}
\end{table}

\noindent \textbf{Dataset Collection.} For this work, we collect precision-register output distributions from a specialized family of Shor-style order-finding circuits run on IBM quantum systems via the \texttt{Qiskit} interface~\cite{qiskit2024}. We restrict to the structured Mersenne / powers-of-two family $N=2^n-1,\; a=2^s \bmod N$. In this case, modular multiplication reduces to cyclic rotations of the work register, and the inverse QFT may be exact or approximate. In our dataset, the two-qubit depth remains below about 2,000. The dataset contains 680 runs collected on five IBM quantum systems. The benchmark instances cover \(N\in\{3,7,15,31,63,127\}\), bases \(a\in\{2,4,8,16\}\), precision sizes \(t\in\{8,10\}\), and a seven-qubit work register. All runs use 4,000 shots and transpilation with \texttt{Qiskit} optimization level 1. Table~\ref{tab:dataset_summary} summarizes the dataset composition. \\

\begin{table*}[t]
\centering
\setlength{\tabcolsep}{2.5pt}
\renewcommand{\arraystretch}{1.10}
\begin{tabularx}{\textwidth}{
    >{\raggedright\arraybackslash}p{0.05\textwidth}
    >{\raggedright\arraybackslash}p{0.20\textwidth}
    >{\raggedright\arraybackslash}p{0.30\textwidth}
    >{\raggedright\arraybackslash}X
}
\toprule
&\textbf{Quantity} & \textbf{Definition} & \textbf{Algorithmic interpretation} \\
\midrule

\multirow{2}{*}{\rotgroup{Distribution\\level}}
&
Autocorrelation peak strength $A_{\mathrm{peak}}$
&
$q(y)=p(y)-u(y)$, \;
$A(\ell)=\sum_y q(y)\,q(y+\ell)$, \;
$A_{\mathrm{peak}}=\max_{\ell\neq 0}A(\ell)/A(0)$
&
\textit{Residual comb spacing.} Larger values indicate that the noisy distribution still preserves repeated-spacing structure associated with order finding. \\
\addlinespace[3pt]

&
Normalized entropy $H_{\mathrm{norm}}$
&
$H_{\mathrm{norm}}(p)=\bigl(-\sum_y p(y)\log p(y)\bigr)/\log Q$
&
\textit{Distribution flatness.} Larger values indicate that probability mass is more broadly spread and less concentrated around order peaks. \\

\cmidrule(lr){1-4}
\addlinespace[2pt]

\multirow{2}{*}{\rotgroup{Post-processing-aware}}
&
Dominant verified mass fraction $M_{1,\mathrm{frac}}$
&
$M_{\mathrm{ver}}=\sum_{r_0} m(r_0)$, \;
$M_1=\max_{r_0} m(r_0)$, \;
$M_{1,\mathrm{frac}}=M_1/M_{\mathrm{ver}}$
&
\textit{Strength of the leading verified denominator after classical post-processing.} Larger values indicate that one candidate order dominates the verified mass. \\
\addlinespace[3pt]

&
Verified margin fraction $\Delta_{\mathrm{ver,frac}}$
&
$M_2=\text{second-largest } m(r_0)$, \;
$\Delta_{\mathrm{ver,frac}}=(M_1-M_2)/M_{\mathrm{ver}}$
&
\textit{Post-processing decision margin}. Larger values indicate clearer separation between the leading verified denominator and the nearest competing verified candidate. \\

\bottomrule
\end{tabularx}
\caption{Distribution-level and post-processing-aware features for noisy precision-register distributions. Here $Q=2^t$ and $u(y)=1/Q$. The verified-candidate masses $m(r_0)$ are defined from the continued-fraction and modular-verification map for the fixed instance $(N,a)$.}
\label{tab:recoverability_features}
\end{table*}

\noindent \textbf{Features for Recoverability.} The ideal output of quantum order finding has a characteristic peak-comb structure. We do not measure agreement with a known peak-comb structure, since that would require prior knowledge of the correct order and of the corresponding comb support. Instead, we investigate features that can be computed from the measured precision-register distribution and from the verified-candidate masses produced by the classical post-processing map, without using the true order in advance. Let $u(y)=1/Q$ denote the uniform distribution. Table~\ref{tab:recoverability_features} summarizes the four quantities used throughout this work. We use these four metrics to separate four distinct questions: \emph{(i)} how flat the distribution is, \emph{(ii)} whether it exhibits repeated peak organization, \emph{(iii)} how strongly the verified probability mass concentrates on the leading candidate denominator, and \emph{(iv)} how well separated that leading verified candidate is from its closest competitor.

The distinction in Table~\ref{tab:recoverability_features} is important for interpreting recoverability. The distribution-level features \(A_{\mathrm{peak}}\) and \(H_{\mathrm{norm}}\) indicate whether the measured distribution still retains order-relevant structure. In contrast, \(M_{1,\mathrm{frac}}\) and \(\Delta_{\mathrm{ver,frac}}\) describe the outcome of the classical decoding map without using the true order \(r\). A distribution can therefore appear noisy at the histogram level but remain recoverable if the verified mass is concentrated on the correct candidate denominator. Conversely, a distribution can retain visible structure but fail if the verified mass favors an incorrect candidate. Thus, the four statistics in Table~\ref{tab:recoverability_features} separate histogram-level structure from decoder-level candidate competition.

\section{Results}
\noindent \textbf{Feature Distribution.} To assess whether recoverability can be anticipated from statistics alone, we compared recoverable and unrecoverable instances using a set of order-agnostic metrics that quantify the amount and organization of residual structure. 

\begin{figure}[t]
    \centering
    \includegraphics[width=\columnwidth]{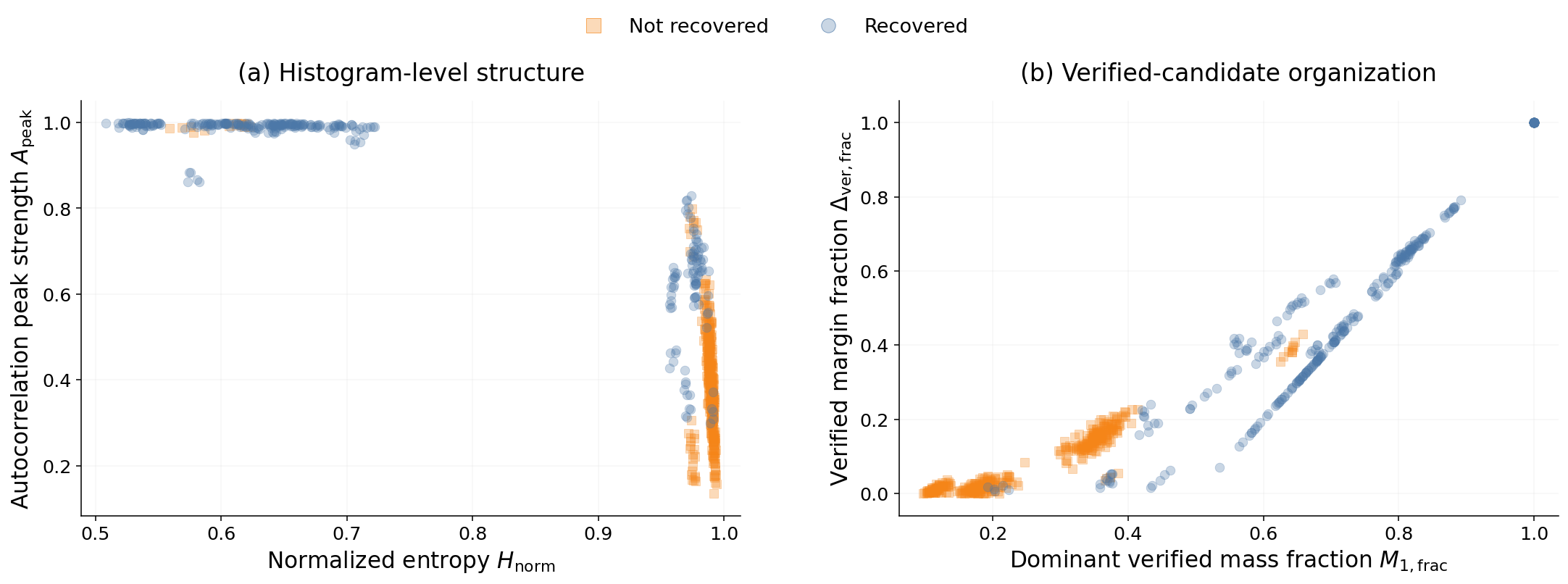}
\caption{Residual structure and recoverability. Each point represents one instance, colored by recoverability outcome. \textbf{(a)} Normalized entropy $H_{\mathrm{norm}}$ versus autocorrelation peak strength \(A_{\operatorname{peak}}\), characterizing flatness and residual structure. \textbf{(b)} Dominant verified mass fraction \(M_{1,\operatorname{frac}}\) versus verified margin fraction \(\Delta_{\operatorname{ver, frac}}\), characterizing the organization of residual structure.}
    \label{fig:residual-structure-recoverability}
\end{figure}

In Figure~\ref{fig:residual-structure-recoverability}(a), recoverable and unrecoverable instances overlap but concentrate in different regions of the entropy–autocorrelation plane. Recoveries are less common in the high entropy, low autocorrelation regime, where the measured distribution is flatter and exhibits weaker repeated-spacing structure. In Figure~\ref{fig:residual-structure-recoverability}(b), recoverable instances tend to occur at higher \(M_{1,\operatorname{frac}}\) and \(\Delta_{\operatorname{ver,frac}}\), indicating that recoverability is favored when one verified denominator dominates and is well separated from its closest competitor. Recoverability therefore depends on both residual structure in the histogram and the organization of probability mass after classical post-processing.

\begin{figure}[t]
    \centering
    \includegraphics[width=\linewidth]{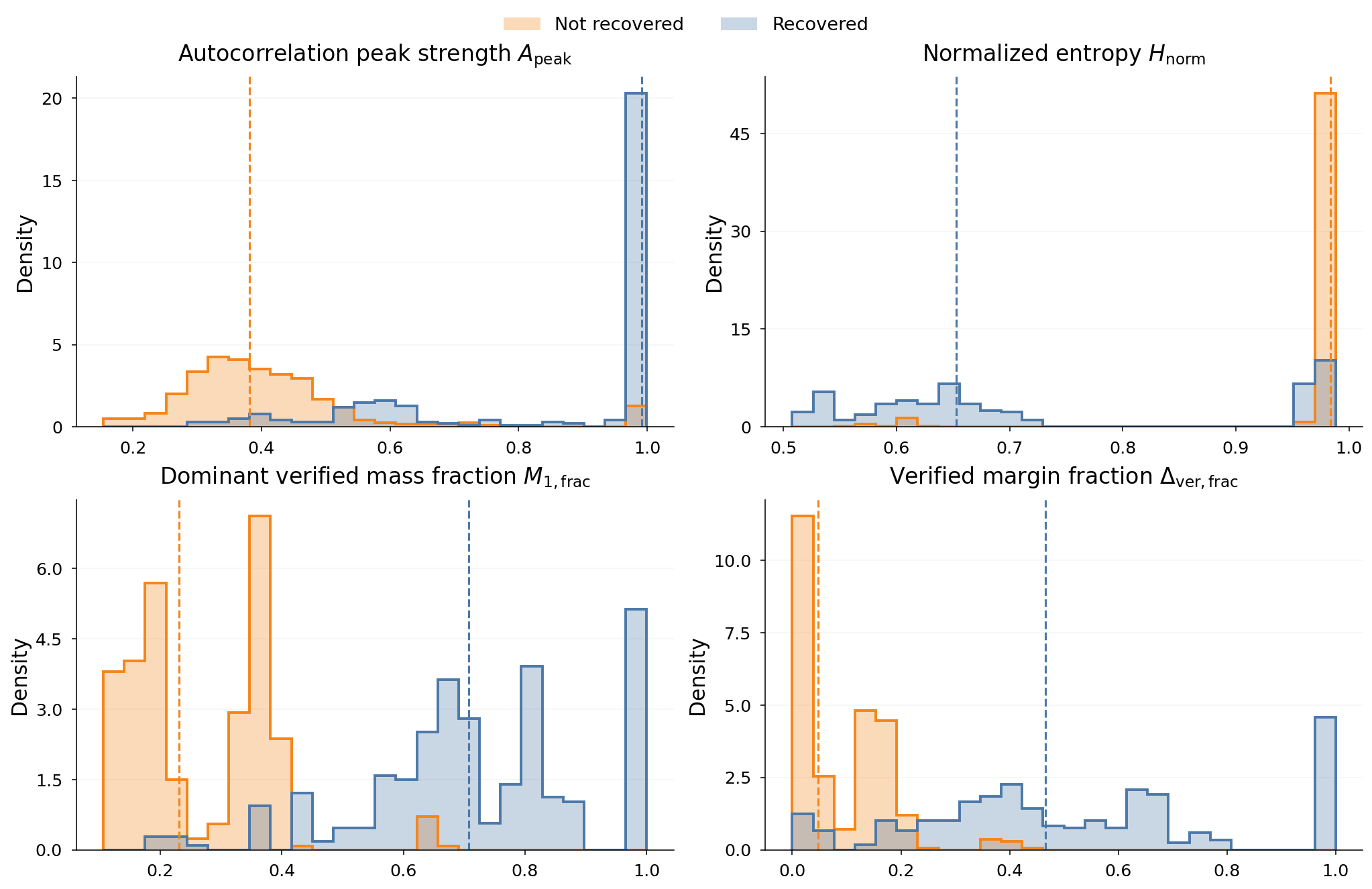}
    \caption{Distribution of four features by recoverability. The histograms compare runs that were not recoverable (orange) with runs that were recoverable (blue) for autocorrelation peak strength $A_{\mathrm{peak}}$, normalized entropy $H_{\mathrm{norm}}$, dominant verified mass fraction \(M_{1,\operatorname{frac}}\) and verified margin fraction \(\Delta_{\operatorname{ver, frac}}\). Bars show empirical density, and dashed vertical lines mark the median of each group.}
    \label{fig:metric-distributions-recoverability}
\end{figure}
Figure~\ref{fig:metric-distributions-recoverability} shows the same trend in the marginal feature distributions. Recovered runs tend to exhibit higher $A_{\mathrm{peak}}$, higher $M_{1,\mathrm{frac}}$, higher $\Delta_{\mathrm{ver,frac}}$, and lower $H_{\mathrm{norm}}$, although the recoverable and non-recoverable classes still overlap substantially. Thus, no single feature reliably separates recoverable from non-recoverable runs on its own. This overlap motivates a closer look at individual cases. The examples in Figure \ref{fig:case1_case2} show that recoverability can be found in highly distorted distributions, and it is not determined by peak presence alone. 

In the top row of Figure~\ref{fig:case1_case2}, the measured histogram is diffuse and its global structure is weak. Nevertheless, the probability mass induced on verified candidate denominators is maximized at the true order, so the instance is recoverable. In the bottom row, the measured distribution retains visible peaks, but the same post-processing map assigns the largest verified mass to an incorrect denominator, so recoverability fails. These examples show that global distribution structure is informative but not sufficient for recoverability. What ultimately matters is how the surviving probability mass is distributed among competing verified candidate denominators. \\

\noindent \textbf{Single-Feature Predictive Power.} To quantify the predictive value of each feature in isolation, we compute the Area Under the Receiver Operating Characteristic curve ($\operatorname{AUROC}$) for recoverability~\cite{fawcett2006introduction}. For a scalar score $s$, $\operatorname{AUROC}$ measures how well the score ranks recoverable runs above non-recoverable runs, independent of any particular threshold. An $\operatorname{AUROC}$ of $1$ indicates perfect separation, while $0.5$ corresponds to random ranking. For quantities whose raw orientation is opposite to recoverability, here the normalized entropy $H_{\mathrm{norm}}$, we reverse the sign so that larger values always correspond to greater predicted recoverability.

\begin{table}[t]
\centering
\setlength{\tabcolsep}{3.5pt}
\renewcommand{\arraystretch}{1.08}
\begin{tabular}{c l c}
\toprule
\textbf{Rank} & \textbf{Metric} & \textbf{AUROC} \\
\midrule
1 & Dominant verified mass fraction $M_{1,\mathrm{frac}}$ & 0.981 \\
2 & Normalized entropy $H_{\mathrm{norm}}$ & 0.946 \\
3 & Verified margin fraction $\Delta_{\mathrm{ver,frac}}$ & 0.946 \\
4 & Autocorrelation peak strength $A_{\mathrm{peak}}$ & 0.937 \\
\bottomrule
\end{tabular}
\caption{Ranking of the four selected features by single-feature AUROC for recoverability.}
\label{tab:auroc_ranking}
\end{table}

Table~\ref{tab:auroc_ranking} reports the single-feature \(\operatorname{AUROC}\) values. Among the predictors, the dominant verified mass fraction \(M_{1,\mathrm{frac}}\) and the verified margin fraction \(\Delta_{\mathrm{ver,frac}}\) are both derived from the distribution of verified candidate denominators and therefore directly reflect how strongly the noisy histogram favors one candidate under classical post-processing. The two global histogram descriptors, autocorrelation peak strength and normalized entropy, are also strong predictors, indicating that both repeated-spacing structure and the overall concentration of probability mass are informative for recoverability. Overall, these results suggest that recoverability is governed most directly by how probability mass is distributed among verified candidates, while global comb-like structure remains informative but less decisive on its own. \\

\noindent \textbf{Multivariate Importance.} Single-feature $\operatorname{AUROC}$ shows how informative each quantity is in isolation, but it does not reveal which quantities still provide distinct information once the others are already known. To address this, we train a random-forest classifier~\cite{breiman2001random} on the four selected features
$A_{\mathrm{peak}}, H_{\mathrm{norm}},M_{1,\mathrm{frac}}, \Delta_{\mathrm{ver,frac}}$, using 5-fold stratified cross-validation. In each split, the model is trained on four folds and evaluated on the held-out fold, with the procedure repeated so that each fold is used once for evaluation. We then assess feature relevance by permutation importance on the held-out data: one feature is randomly permuted at a time, and the resulting decrease in $\operatorname{AUROC}$ is recorded. A larger $\operatorname{AUROC}$ drop means that the classifier relies more strongly on that feature once the full feature set is taken into account. This random forest permutation analysis is the primary quantitative feature-importance measure in this section.

\begin{figure*}[t]
    \centering
    \includegraphics[width=0.9\linewidth]{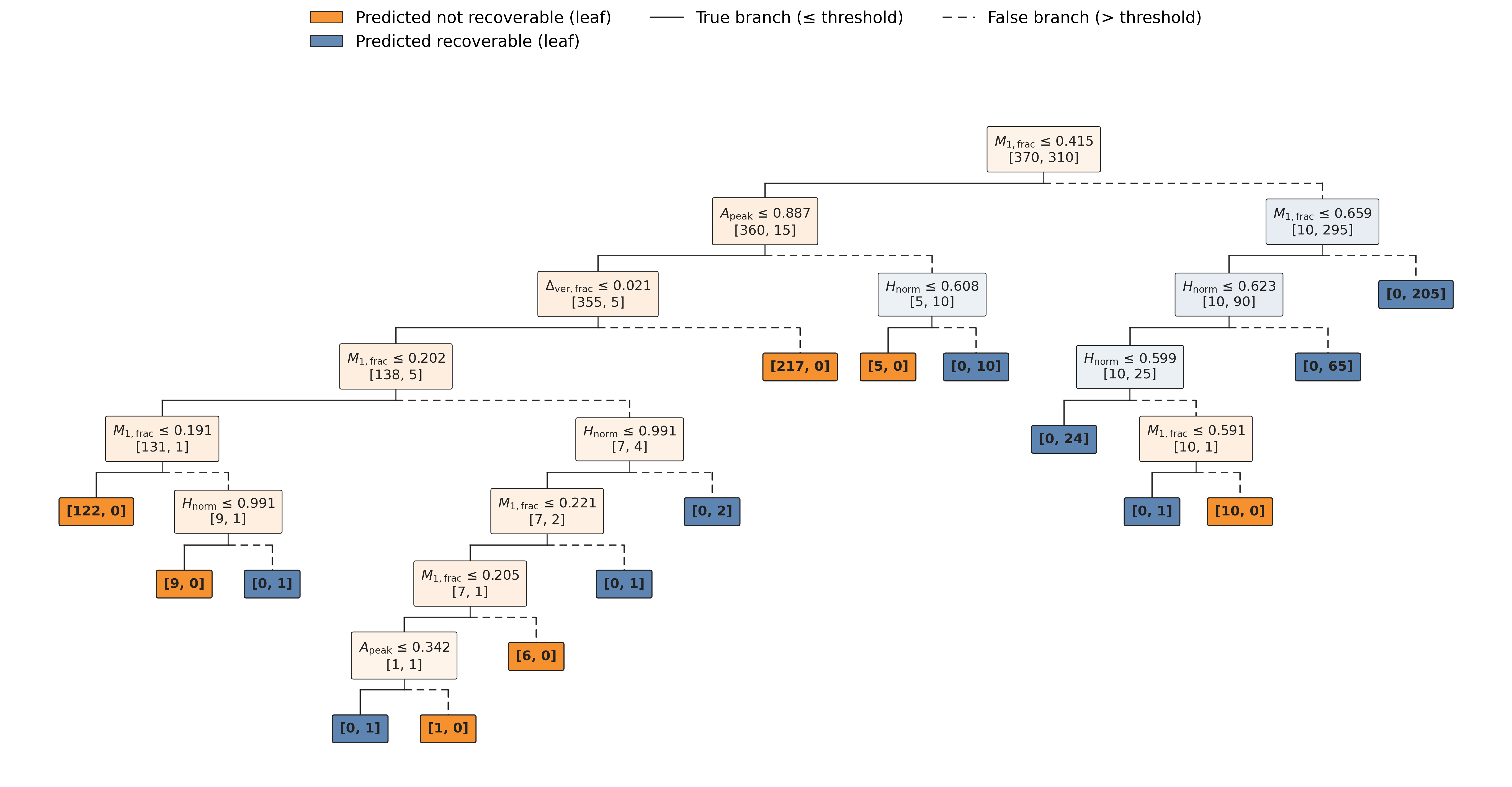}
   \caption{Decision tree classifier for predicting recoverability. Internal nodes show split rules based on  $A_{\rm peak}$, $H_{\rm norm}$, $M_{1,\rm frac}$, and $\Delta_{\rm ver,frac}$.  Terminal leaves show the final prediction, with orange indicating predicted not recoverable  and blue indicating predicted recoverable. Counts are shown as  $[\text{not recoverable},\text{ recoverable}]$. Solid black branches indicate the true  condition, $\leq$ threshold, while dashed black branches indicate the false condition,  $>$ threshold.}
    \label{fig:tree_diagram}
\end{figure*}

\begin{figure}[t]
    \centering
    \includegraphics[width=\columnwidth]{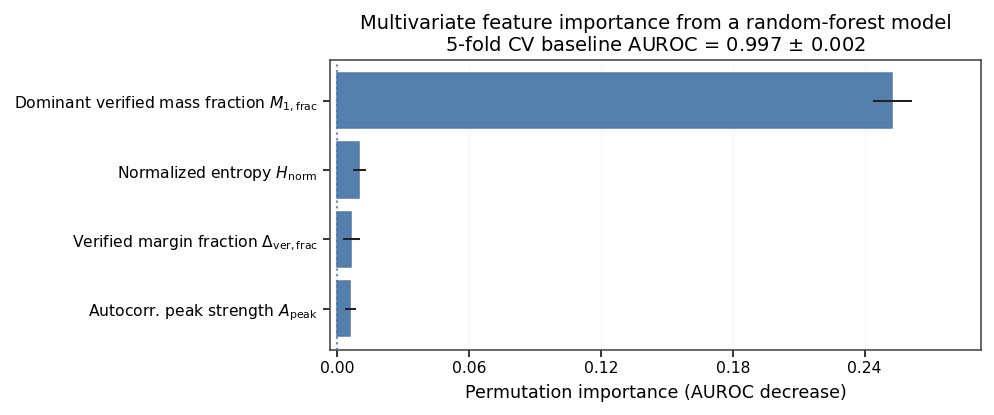}
    \caption{Permutation importance from a random-forest classifier trained on the four selected recoverability features. Importance is measured as the decrease in held-out $\operatorname{AUROC}$ after randomly permuting each feature. Error bars show the standard deviation across cross-validation folds.}
    \label{fig:perm_importance}
\end{figure}

Figure~\ref{fig:perm_importance} summarizes the resulting multivariate importance values. The overall predictive performance is high, with $\operatorname{AUROC}$ $=0.997\pm0.002$, indicating that these four quantities together almost perfectly separate recoverable from non-recoverable runs in this dataset. By a clear margin, the dominant feature is the verified mass fraction $M_{1,\mathrm{frac}}$. The remaining features have much smaller permutation importance once $M_{1,\mathrm{frac}}$ is included. In the cross-validated random-forest analysis, \(H_{\operatorname{norm}}\) gives the largest residual contribution among the secondary features, while $\Delta_{\mathrm{ver,frac}}$ and $A_{\operatorname{peak}}$ have smaller effects on held-out \(\operatorname{AUROC}\).

The strong role of $M_{1,\mathrm{frac}}$ shows that recoverability is governed most directly by how strongly the noisy histogram concentrates verified probability mass on a single candidate denominator. The smaller importance of \(H_{\operatorname{norm}}\), \(\Delta_{\operatorname{ver,frac}}\) and \(A_{\operatorname{peak}}\) does not mean that these quantities are irrelevant in every case. Instead, it indicates that they contribute little additional information in this dataset once the dominant verified-mass feature is known. In particular, \(\Delta_{\operatorname{ver,frac}}\) overlaps substantially with $M_{1,\mathrm{frac}}$, while \(A_{\operatorname{peak}}\) and $H_{\mathrm{norm}}$ can still help refine localized boundary cases without strongly affecting the global held-out \(\operatorname{AUROC}\). \\

\noindent \textbf{Interpretable Threshold Rules.} To complement the random-forest analysis with an interpretable rule-based view, Figure~\ref{fig:tree_diagram} shows a single decision tree~\cite{breiman2017classification} trained on the same four features. This tree provides an explicit threshold-based description of how recoverable and non-recoverable runs are partitioned. Each internal node applies a threshold to one metric, and each path defines a sequence of conditions leading to a terminal prediction. The bracketed values show the class counts as \([\text{not recoverable},\text{ recoverable}]\) for the runs reaching that node.

A few qualitative conclusions emerge from analyzing the decision tree. First, the root split is on the dominant verified mass fraction \(M_{1,\mathrm{frac}}\), confirming that this is the primary quantity controlling recoverability. The threshold \(M_{1,\mathrm{frac}}\le 0.415\) already separates the dataset into a mostly non-recoverable branch and a mostly recoverable branch, while \(M_{1,\mathrm{frac}}>0.659\) leads to a pure recoverable leaf. This indicates that once the verified probability mass is strongly concentrated on a single candidate denominator, recoverability becomes almost deterministic in this dataset. On the non-recoverable side, very small \(M_{1,\mathrm{frac}}\) is itself almost sufficient for failure, while the secondary splits on \(A_{\mathrm{peak}}\), \(H_{\mathrm{norm}}\), and \(\Delta_{\mathrm{ver,frac}}\) refine the intermediate and boundary cases. 

\section{Discussion and Conclusion}

This work studied quantum noise in order finding at the application level, with the measured precision-register distribution as the main object of analysis. We showed that noisy order finding remains recoverable when the measured distribution preserves enough order-relevant structure for classical post-processing to favor the true verified order over competing candidates. Failure occurs when noise redistributes probability mass toward incorrect verified denominators, even if the distribution still appears visibly structured. Thus, visual structure alone is not sufficient; the relevant question is whether the remaining structure is aligned with the continued-fraction and verification steps used to recover the order.

Global, order-agnostic metrics such as normalized entropy and autocorrelation peak strength explain a substantial part of recoverability but do not fully determine it. These metrics capture broad properties of the distribution, such as concentration, flatness, and residual periodicity, but they do not directly encode how probability mass is distributed among candidate orders after continued-fraction decoding and modular verification. As a result, two distributions with similar entropy or autocorrelation behavior can lead to different recovery outcomes. Circuit depth is an obvious feature in this setting, since it captures the overall circuit scale and exposure to noise; however, it is ultimately a coarse proxy and becomes less informative in borderline cases.

The post-processing-aware quantities provide a more direct view of the recovery mechanism. In particular, the dominant verified mass fraction $M_{1,\mathrm{frac}}$ measures how strongly the verified probability mass concentrates on the leading candidate denominator, while the verified margin fraction $\Delta_{\mathrm{ver,frac}}$ measures how well separated that leading candidate is from its closest competitor. These quantities do not require prior knowledge of the true order, but they are aligned with the same continued-fraction and modular-verification map used by the classical decoder. The strong role of $M_{1,\mathrm{frac}}$ therefore shows that recoverability is governed most directly by whether the noisy histogram gives the post-processing routine a dominant verified candidate. The global metrics, especially $A_{\mathrm{peak}}$ and $H_{\mathrm{norm}}$, remain useful for describing residual structure in the measured distribution, but their relevance is mediated by how that structure maps onto verified candidate denominators.

An interpretation of the observed noisy precision-register distributions, such as those in Figure~\ref{fig:case1_case2}, is as a \emph{two-stage quantum noise propagation process}. In the first stage, errors in the modular exponentiation block perturb the ideal action of the Shor circuit before the inverse QFT is applied. For the approximation used in our data collection  (Mersenne / powers-of-two family with \(N=2^n-1\) and \(a=2^s \bmod N\)), the underlying modular multiplication reduces to a cyclic rotation of the work register. The ideal circuit can therefore be associated with a single effective sector determined by the intended shift parameter \(s\). We use the term \emph{sector} to denote an effective modular-multiplication action close to this intended one. Noise or compilation error can populate nearby sectors, which we index by \(h\in\mathcal{H}\), with the intended sector given by \(h=s\). After phase estimation, these sectors appear as distinct comb families in the precision register. For more general modular exponentiation, however, this sector picture becomes less natural, because the work-register unitary is no longer a simple cyclic rotation but a more complicated arithmetic permutation. In that case, noise may produce more irregular precision-register distortions, and a small mixture of nearby comb families may no longer provide an adequate description. In the second stage, noise in the inverse QFT, together with measurement noise and other late-stage errors, broaden and deform these comb families in the measured histogram.

A simple application-level model of this noise propagation is
\begin{equation}
\begin{aligned}
p(y)\approx\;&
\underbrace{(1-\varepsilon)\,[\,p_s * K_{\sigma_0}\,](y)}_{\text{broadened intended comb family}} \\
&+\;
\underbrace{\varepsilon \sum_{h\in\mathcal{H}\setminus\{s\}} \nu_h \,[\,p_h * K_{\sigma_h}\,](y)}_{\text{noise-induced competing comb families}},
\qquad
\sum_{h\in\mathcal{H}\setminus\{s\}} \nu_h = 1,
\end{aligned}
\end{equation}
where \(p_s\) is the ideal comb family associated with the intended sector \(h=s\), the \(p_h\) with \(h\neq s\) are competing comb families populated by noise or compilation error, \(\varepsilon\in[0,1]\) is the total weight transferred out of the intended family, and the coefficients \(\nu_h\) distribute that leaked weight among the competing families. The kernels \(K_{\sigma_0}\) and \(K_{\sigma_h}\) are normalized broadening kernels, with \(*\) denoting convolution over the precision-register index. In this picture, the dominant term \((1-\varepsilon)[\,p_s * K_{\sigma_0}\,](y)\) represents the broadened ideal comb family produced by the intended modular action, while the smaller leakage term redistributes probability mass into competing comb families. Accordingly, jump-like structures reflect noise-induced transfer between comb families. Diffuse broadening reflects local deformation of each family due to imperfections in the inverse QFT and other late-stage noise.

An important direction for future work is to expand the dataset so that it covers a larger and more continuous region of the distributional feature space. In the current setup, some regions remain unpopulated, for example, the observed gap in the normalized-entropy range. Filling such gaps, if possible, would allow a more precise characterization of the transition between recoverable and non-recoverable regimes, and would improve the robustness of downstream predictive or mitigation models. A broader dataset would also help determine which recoverability indicators are universal across problem sizes, circuit families, backend noise profiles, and approximation levels, and which are specific to the present hardware regime.

These results motivate an application-level approach to quantum noise and error mitigation. A direction for future work is to design mitigation methods targeted directly at order recovery from noisy distributions by modifying or denoising the measured distribution so as to better preserve the dominance of the true verified order. The decision-tree analysis also suggests a possible route toward application-specific quantum error mitigation. For instance, mitigation for noisy order finding should aim to shift runs across the interpretable recoverability thresholds identified by the tree, for example by increasing the dominant verified mass fraction, reducing entropy, or strengthening residual comb-like structure.

\begin{acks}
The authors gratefully acknowledge the Wallenberg Centre for Quantum Technology (WACQT) for access to the IBM Quantum Computing systems used in this work.
\end{acks}

\bibliographystyle{ACM-Reference-Format}
\bibliography{references}

\end{document}